\documentclass[lettersize,journal]{IEEEtran}
\usepackage{amsmath,amsfonts}
\usepackage{algorithmic}
\usepackage{algorithm}
\usepackage{array}
\usepackage[caption=false,font=normalsize,labelfont=sf,textfont=sf]{subfig}
\usepackage{textcomp}
\usepackage{stfloats}
\usepackage{xcolor}
\usepackage{url}
\usepackage{verbatim}
\usepackage{graphicx}
\usepackage{cite}
\usepackage[utf8]{inputenc}
\usepackage[english]{babel}
\usepackage{fancyhdr}

\begin{document}

\title{Automatic Monitoring of Fruit Ripening Rooms by UHF RFID Sensor Network and Machine Learning}

\author{Cecilia Occhiuzzi\textsuperscript{$\ast \$$}, Francesca Camera\textsuperscript{$\$$}, Michele D’Orazio\textsuperscript{£}, Nicola D’Uva\textsuperscript{$\$$}, Sara Amendola\textsuperscript{$\$$}, Giulio Maria Bianco\textsuperscript{$\ast\$$}, Carolina Miozzi\textsuperscript{$\$$}, Luigi Garavaglia$^{\circ}$, Eugenio Martinelli\textsuperscript{£}, Gaetano Marrocco\textsuperscript{$\ast\$$}

\begin{center}
{\small $\ast$ University of Roma ``Tor Vergata", DICII, Via del Politecnico, 1- 00133 Rome, Italy}\\
{\small \textsuperscript{$\$$}RADIO6ENSE srl, Via del Politecnico 1- 00133 Rome, Italy}\\
\textsuperscript{£ }{\small University of Roma ``Tor Vergata", DIE, Via del Politecnico, 1- 00133 Rome, Italy}\\
{\small $^{\circ}$ ILPA Group S.p.A, Via Castelfranco, 52 - 40053 Valsamoggia (BO) - Italy}
\end{center}

}

\markboth{Journal of \LaTeX\ Class Files,~Vol.~14, No.~8, August~2021}%
{Shell \MakeLowercase{\textit{et al.}}: A Sample Article Using IEEEtran.cls for IEEE Journals}

\IEEEpubid{0000--0000/00\$00.00~\copyright~2021 IEEE}

\maketitle

\begin{abstract}
Accelerated ripening through the exposure of fruits to controlled environmental conditions and gases is nowadays one of the most assessed food technologies, especially for climacteric and exotic products. However, a fine granularity control of the process and consequently of the quality of the goods is still missing, so the management of the \textit{ripening rooms} is mainly based on qualitative estimations only. Following the modern paradigms of Industry 4.0, this contribution proposes a non-destructive RFID-based system for the automatic evaluation of the live ripening of avocados. The system, coupled with a properly trained automatic classification algorithm based on Support Vector Machines (SVMs), can discriminate the stage of ripening with an accuracy greater than 85$\%$. 
\end{abstract}


\section{Introduction}

\IEEEPARstart{F}{ruits}, vegetables, cheeses, and cured meats are usually artificially ripened in ripening rooms \cite{harvey1928artificial}. Especially for fruits and vegetables, modern equipment is designed to modify and fine-tune specific environmental conditions, such as temperature, relative humidity, ethylene, carbon dioxide \(\left(CO_{2}\right)\)and oxygen concentration \(\left(O_{2}\right)\). The final aim is to produce goods whose ripening status is suitable for the different steps of the distribution chain, from \textit{stock} to \textit{final customer}.\par
To control the evolution of the process and hence the status of the fruits, the firmness of the pulp is usually evaluated. The monitoring is frequently done by removing samples from the room and by performing tests with penetrometers or durometers \cite{magwaza2015a}. Often, qualified operators manually evaluate the effectiveness of the process by visual inspection, palpation, and tasting \cite{pamungkas2020evaluation}. All these methodologies are usually destructive and qualitative and are not suitable for continuous implementations. Furthermore, the recurring opening and closing of the ripening room alters the internal environment leading to a waste of time, energy and, in general, of quality control \cite{no_auth_1}. Accordingly, the non-destructive real-time quality inspection would be vital. Computer vision, potentiometric, and ultrasonic techniques have been already proposed, as well as methods involving the sampling of the gases surrounding the fruits \cite{bhargava2021fruits, Verma15, Corrieu18, soltani2011evaluating}. However, the accuracy and cost-effectiveness of the most innovative systems are not compliant with typical industrial demands in terms of costs, easiness of implementation and speed of the analysis so that the problem is still open and represents a practical challenge for producers, importers, and fruit distributors \cite{hussain2018innovative}.\par 
{\footnotesize \begin{figure}[tp]
\includegraphics[width=8.58cm,height=5.93cm]{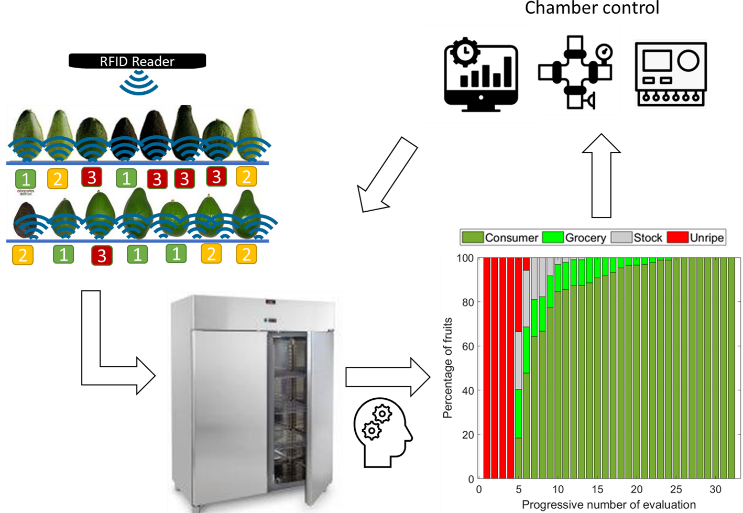}
\caption{Scheme of the proposed monitoring system. Inside the ripening chamber, an RFID reader continuously interrogates passive tags in contact with fruits. By processing the retrieved data through Artificial Intelligence algorithms, the ripening status could be estimated, and the chamber operative conditions could be consequently adapted in terms of exposure time, ethylene concentration, RH (relative humidity) and temperature.}
\label{fig:fig_scheme_proposed_monitoring_system}
\end{figure}}
In recent years, a particular focus has been devoted to avocado ripening due to this fruit's increased usage and demand in the food and cosmetic industry \cite{pinto2019classification, villasenormora2017correlation}. Being a climacteric fruit, the avocado requires artificial ripening too. Namely, to preserve a firm pulp texture, good color, and flavor while avoiding contusion, avocados are cut from the plant at a mature-green stage and exported to consumer countries. The imported avocado fruits are kept in airtight warehouses with a temperature and an ethylene gas control system until their ripening stage is suitable for distribution \cite{mizrach2000monitoring}. The time required for ripening avocado varies according to the ripeness of the fruit when it is received and depends on the cultivation and harvesting conditions. Furthermore, each importer/distributor has its ripening processes and procedures \cite{babiker2020influence}. Ripening is hence highly customized and variable across the distribution chain.\par
Various chemical-physical indicators, data carriers and sensors have been explored for monitoring avocados' quality and ripening, mainly at the packaging level \cite{ghaani2016an}. Most of them are related to the freshness and the integrity of the packaging and are intended for temperature, leakage, chemical-based, (pH, humidity, gas sensor), and visual appearance applications \cite{Firouz21}. Limitations to their spread acceptance are again related to costs, analysis easiness, and response robustness. In particular, visual techniques, despite the limited cost and the easiness in implementation, suffer the dependence of the color on the harvest conditions, the storage temperature, and the fruit variety. Hence they are mainly limited to qualitative evaluations for final customers.\par
Thanks to the current advances in RFID-based automatic monitoring of objects \cite{Bartoletti18} and even of people \cite{Bianco22}, the authors recently proposed the use of RFID technology for monitoring the ripening of avocado packaged for final customer distribution in stores \cite{occhiuzzi2020radio}. The system comprises intelligent packaging integrating a single passive UHF RFID tag and a customized reader. It can retrieve the state of the fruit and classify it as \textit{unripe}, \textit{ripe} or \textit{overripe} through a decision tree classification algorithm. The monitoring platform is, however, limited to a single avocado per time, and hence it is not suitable for the application to a multitude of fruits as in an industrial maturation chamber.\par
This paper now proposes a multi-fruit multi-tag UHF ($860$-$960$ MHz), passive, non-destructive RFID system for monitoring avocados' state in an industrial environment. A trolley hosting 128 fruits and integrating an interrogation RFID network with multiple reader antennas is presented. Unlike the previous works, ripening is evaluated by means of three tags for each fruit properly disposed on the basal region so that the detection capability is improved and made more effective. By exploiting the \textit{sensor-less} approach in \cite{Occhiuzzi13}, signals received and backscattered during communications are exploited to retrieve indirect information about the status of the fruits. Up to eight interrogation modalities are implemented to collect different power and phase signals from the tags. Electromagnetic indicators are then used to feed a classification algorithm based on a Support Vector Machine (SVM) which is capable of discriminating the state of the fruits among four possible classes, corresponding to different ripening stages for different storages, from stock to consumer distributions. Monitoring is periodic; namely, each fruit is continuously sampled and evaluated up to four times in a single day. In agreement with the modern Industry 4.0 trend, the system is aimed at supporting and finally controlling the environmental conditions of the room (time, temperature, relative humidity RH, concentrations of ethylene, \(CO_{2}\) and \( O_{2}\)) depending on the state of the monitored fruits (functioning scheme in Fig.~\ref{fig:fig_scheme_proposed_monitoring_system}). \par
The paper is organized as follows. The rationale of the monitoring approach is briefly recalled, together with the adopted interrogation modalities in Section~\ref{sec:Rationale}. In Section~~\ref{sec:Trolley}, the Smart Trolley is introduced concerning hardware and software components. The prototype and the implementation are then described in Section~~\ref{sec:Implementation}. Section~\ref{sec:Classification} focuses on the classification algorithm and the achieved results, while the error analysis and the performance assessment are finally presented in Section~~\ref{sec:Error}.
\section{Rationale}\label{sec:Rationale}
\begin{figure}[tp]
\centering
\includegraphics[width=6.05cm,height=4.69cm]{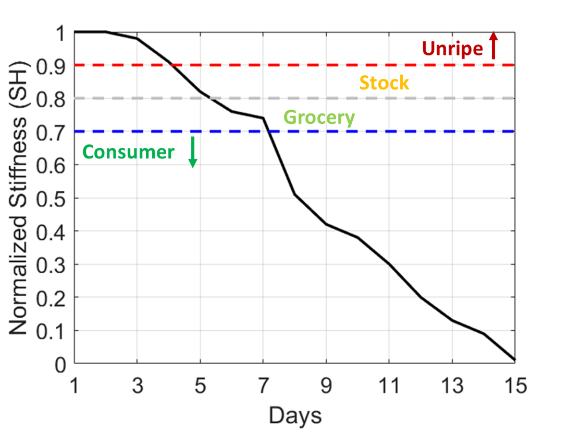}
\caption{Example of SH versus ripening days.}
\label{fig:fig_example_sh_ripening_days}
\end{figure}

During ripening, avocado fruits undergo several chemical and physical variations of both peel and flesh \cite{Schaffer13, Peleg89}. Macroscopically, such variations produce a progressive softening of the fruit, which is generally sensed from the peel as the Shore value (SH), i.e., the firmness and hardness of the fruit as measured by a durometer. Normalizing the SH by the initial value at the beginning of the ripening process, a typical monotonic decreasing profile can be retrieved (an example is visible in Fig.~\ref{fig:fig_example_sh_ripening_days}), whose slope is randomly affected by the fruit variability, the harvest conditions, and by the environmental parameters during the storage. According to the SH levels, four classes (C) of ripening can be considered \cite{occhiuzzi2020radio} based on the thresholds \(\left\{ TH_{n}\right\} =\left\{ 0.9, 0.8, 0.7\right\}\) (Table~\ref{tab:table_i}). Typically, at ambient temperature and without forced environmental conditions, the fruits reach the last ripening class in $5$-$7$ days. In the case of artificial ripening rooms, the previous interval can be further reduced to $3$-$5$ days.
\begin{table}[tp]
\begin{center}\caption{Considered Ripening Classes}
\begin{tabular}{|l|l|}
\hline
\multicolumn{1}{|p{3.5cm}}{\centering
\textbf{CLASS}} & 
\multicolumn{1}{|p{3.5cm}|}{\centering
\textbf{CONDITION}} \\ 
\hline
\multicolumn{1}{|p{3.5cm}}{C1 (unripe)} & 
\multicolumn{1}{|p{3.5cm}|}{\raggedright
SH\( \geq\)0.9} \\ 
\hline
\multicolumn{1}{|p{3.5cm}}{C2 (stock)} & 
\multicolumn{1}{|p{3.5cm}|}{\raggedright
0.8\( \leq\)SH\( <\)0.9} \\ 
\hline
\multicolumn{1}{|p{3.5cm}}{C3 (grocery)} & 
\multicolumn{1}{|p{3.5cm}|}{\raggedright
0.7\( \leq\)SH\( <\)0.8} \\ 
\hline
\multicolumn{1}{|p{3.5cm}}{C4 (consumer)} & 
\multicolumn{1}{|p{3.5cm}|}{\raggedright
SH\( <\)0.7} \\ 
\hline
\end{tabular}
\label{tab:table_i}
\end{center}
\end{table}
The chemical and physical modifications of the fruit reproduce in a macroscopic way the variation of the RF dielectric properties, namely the dielectric permittivity and conductivity \(\left\{ \varepsilon_{r},\sigma [\frac{S}{m}]\right\}\). If an antenna is placed in close proximity of the fruit (e.g., directly attached to the peel), any variation of the condition $Y\left(t\right)$ will, in turn, produce modifications of the antenna impedance $Z\left[Y\left(t\right)\right]$ and of the radiation gain $G\left[Y\left(t\right)\right]$, and definitively of the received and transmitted signals. By exploiting and mastering this phenomenon, a self-sensing completely sensor-less non-destructive passive test system is therefore obtained, wherein the sensor is the antenna, and the antenna is the sensor \cite{Occhiuzzi13}. \par
In a typical UHF RFID platform, the following electromagnetic indicators affected by the fruit condition are returned by the reader.
\begin{itemize}
	\item $P_{in }^{to}\left[\Psi\left(t\right)\right]$ - The turn-on power (dBm): the minimum power entering the reader antenna to power up the tag.
	\item $RSSI\left[\Psi(t)\right]$ - The received signal strength indicator (dBm) related to the backscattered power \( P_{R\leftarrow T }^{}[\Psi\left(t\right)]\) from the tag to the reader.
	\item $\Phi\left[\Psi\left(t\right)\right]$ - The phase of the backscattered signals (deg).
\end{itemize}
Additional metrics can be then derived to define the electromagnetic fingerprint of the fruit. The problem is, then, estimating the ripening process $\Psi\left(t\right)$ through calibration curves or artificial intelligence algorithms capable of recognizing ripening patterns and states.
It is worth noticing that the considered sensing approach is completely unspecific (any variation of the communication link, even not related to a modification of the fruit, can alter the measured RF indicators). Robust control of the setup in terms of reader-tag-fruits mutual positions, as well as robust interrogation and processing algorithms, are required since the sensing is entirely analog and, hence, it is prone to be affected by undesired disturbs of the channels \cite{occhiuzzi2016precision}. 
\section{Smart Trolley}\label{sec:Trolley}
The proposed monitoring system is a 4-shelves trolley (scheme in Fig.~\ref{fig:fig_a_trolley_automatic_monitoring}) capable of simultaneously evaluating the ripening of multiple avocados. The system operates in the UHF band from $860$ to $960$ MHz, thus at both ETSI and FCC frequencies. Fruits are interrogated through a multiple-antenna network, controlled by a custom software operating on a laptop. 
\subsection{Reading Architecture}
\begin{figure}[tp]
\centering
\includegraphics[width=8.4cm,height=5.32cm]{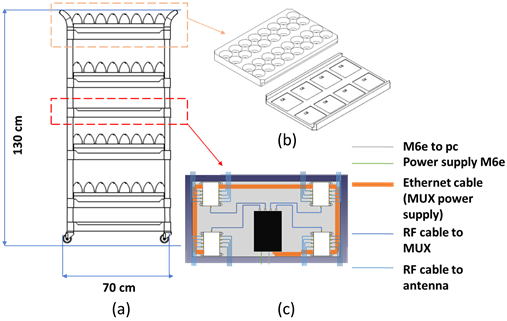}
\caption{(a) Trolley for automatic monitoring of the ripening of avocados into ripening rooms. Four shelves host 128 fruits (32 on each shelf), each in a proper polyurethane lodging. (b) Underneath the fruit lodging, there is the antenna lodging hosting eight near-field patch antennas. (c) All the electronic components (reader and multiplexers) are hosted in a separate lodge.}
\label{fig:fig_a_trolley_automatic_monitoring}
\end{figure}

\begin{figure}[tp]
\centering
\includegraphics[width=6.35cm,height=7.55cm]{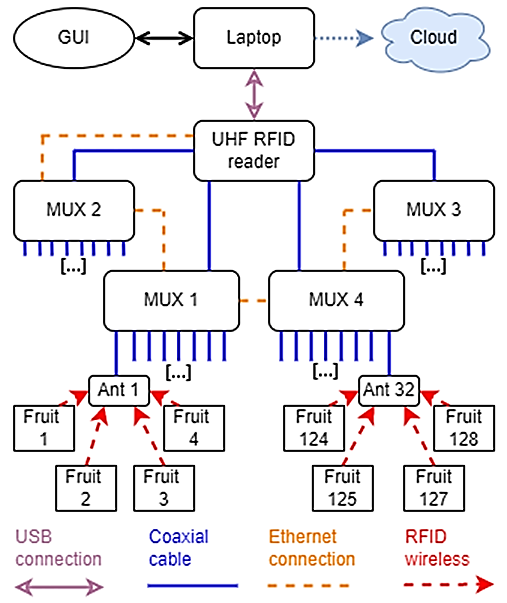}
\caption{Schematic multi-level architecture of the RFID Trolley.}
\label{fig:fig_schematic_multilevel_architecture_rfid}
\end{figure}

The trolley hosts a single M6 ThingMagic reader that, through 4 AdvanMux-8 multiplexers (MUX) by Keonn (see Fig.~\ref{fig:fig_a_trolley_automatic_monitoring}c), controls $32$ near-field antennas (architecture scheme in Fig.~\ref{fig:fig_schematic_multilevel_architecture_rfid}). Each shelf houses (in the \textit{antenna lodging}) eight radiating elements overall capable of monitoring $32$ avocados (see Fig.~\ref{fig:fig_a_trolley_automatic_monitoring}b). The antenna topology and positioning were chosen to guarantee an almost uniform and localized field coverage for the overhead avocados. Each antenna interrogates four fruits through the planar $12$~cm~x~$12$~cm Advantenna-L11 by Keonn.
\subsection{Sensing Tags}
\begin{figure}[tp]
\centering
\includegraphics[width=6.13cm,height=6.14cm]{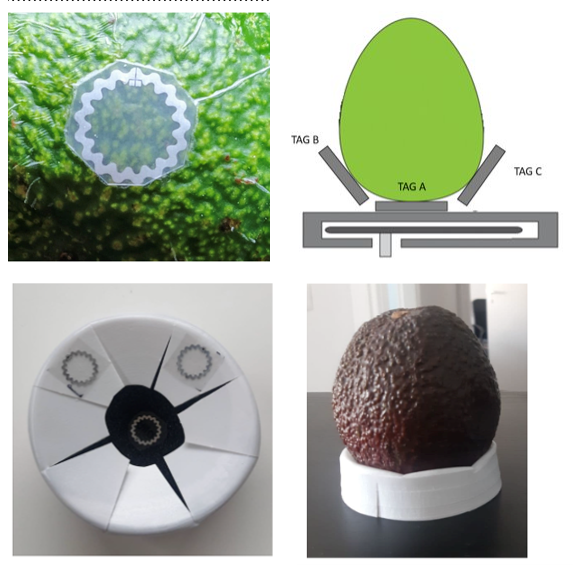}
\caption{Sensing tags for monitoring the ripening of the avocado. A tri-leaf shaped plastic support hosts the three tags and guarantees a firm position of the fruit all along the monitoring period. Tags are in contact with the fruit in three points.}
\label{fig:fig_sensing_tags_monitoring_ripening}
\end{figure}

For each shelf, fruits are hosted in circular cavities carved into the polyurethane \textit{fruit lodging} (see Fig.~\ref{fig:fig_a_trolley_automatic_monitoring}b). The size of the cavities guarantees a vertical positioning of the fruits. Tags are integral with the shelf, three for fruit, and are directly attached on a multi-leaf shaped plastic support to be inserted into the circular cavity (Fig.~\ref{fig:fig_sensing_tags_monitoring_ripening}). The support is designed to allow perfect contact between fruit and tags regardless of variety and size and to further strengthen the placement of the fruit in the cavity. In this way, the position between tags and antennas remains stable during all the days of measurements. Typical signal fluctuations caused by variable setups \cite{occhiuzzi2016precision} are hence sensibly reduced. Finally, being the tags integral with the shelf, the trolley can be quickly loaded and unloaded, with minimal impact on the factory processes. \par
The adopted tag is a circular loop integrating the IC Monza R6-P by Impinji, capable of guaranteeing a good trade-off between communication and sensing features. More  specifically, through the auto-tuning AT modality, the IC can automatically adjust the imaginary part of its input impedance according to the tag operating conditions to optimize the power harvest capability continuously. Alternatively, through reader programming, IC impedance can be kept fixed (AT-off). \par
Tags are placed so that they are in contact with the fruit in the basal region, one directly on the basis (Tag A) and two slightly eccentric toward the equatorial region (Tag B and C), as shown in Fig.~\ref{fig:fig_sensing_tags_monitoring_ripening}. Being the ripening process sensibly variable among fruits in terms of starting point and evolution, the possibility to have three tags for each fruit enables a triple and a spatially distributed sampling, with benefits in terms of monitoring performances (early detection of the ripening) and robustness of the algorithm (processing of multiple independent signals). \par
\subsection{Software Architecture}
The trolley is governed by a software module written in C$\#$ that enables a multi-level control of the monitoring, providing the system with the possibility of dynamically customizing the activity in terms of:
\begin{itemize}
	\item fruit selection (the part of the trolley to be monitored), and
	\item mode of operation (i.e., the interrogation modality).
\end{itemize}
The reader periodically interrogates up to $128\times 3 = 384$ tags. Each tag, according to a proper \textit{tags list} based on the known position of the tags, interacts only with a specific reader antenna, therefore avoiding multiple readings by adjacent elements and hence errors in retrieving power and phase indicators. The interrogation procedure runs as follows (see the architectural scheme in Fig.~\ref{fig:fig_schematic_multilevel_architecture_rfid} again):
\begin{enumerate}
	\item selection of the multiplexer MUX\textsubscript{i }with \( i =\left\{ 1,\ldots ,4\right\}\);
	\item selection of the single antenna \( A_{i,j}\) by activating the corresponding port \( j =\left\{ 1,\ldots ,8\right\}\);
	\item selection of the fruit by choosing the tags \( T_{i,j,k,t}\)\textsubscript{ }with \( k =\left\{ 1,2,3,4\right\}\) indicating the fruit and \( t =\left\{ A,B,C\right\}\) denoting the position of the tag.
\end{enumerate}
The previously described electromagnetic indicators $\{$$P_{in}^{to},RSSI,\Phi$$\}$ are retrieved by imposing four interrogation modalities involving different settings of the interrogation frequencies $f$ and of the power entering into the reader antenna $P_{in}$. More specifically, the reader interrogates the tags by swapping:
\begin{enumerate}
	\item the power entering into the reader antenna $P_{in}$ in a defined range \(\left[P_{in }^{\min }, P_{in }^{Max}\right]\), in the ETSI band ($865$-$868$ MHz);
	\item the power entering into the reader antenna \( P_{in}\) in a defined range, \(\left[P_{in}^{\min}, P_{in}^{Max}\right]\), in the FCC band ($903$-$928$ MHz);
	\item the interrogation frequency in ETSI and FCC bands for \(P_{in}=30\) dBm;
	\item the interrogation frequency in the ETSI and FCC bands to evaluate the turn-on power \(P_{in}^{to}\). 
\end{enumerate}
Since the tags integrate new-generation auto-tuning ICs, the previous RF indicators can be measured with the auto-tuning capability enabled (AT on) or, instead, disabled (AT off) to provide additional indicators. Consequently, the eight interrogation modalities are exploited to retrieve the data to feed the algorithm. Depending on the selected power ranges \(\left[P_{in}^{\min}, P_{in}^{Max}\right],\) the reader takes up to $35$ seconds to complete the entire interrogation of a single tag and, overall, up to $4$ hours to scan all the $128$ fruits. \par
Data retrieved during RF interrogation is then processed through machine learning techniques, later on described.
\section{Smart Trolley: Implementation}\label{sec:Implementation}
\begin{figure}[tp]
\centering
\includegraphics[width=8.63cm,height=5.52cm]{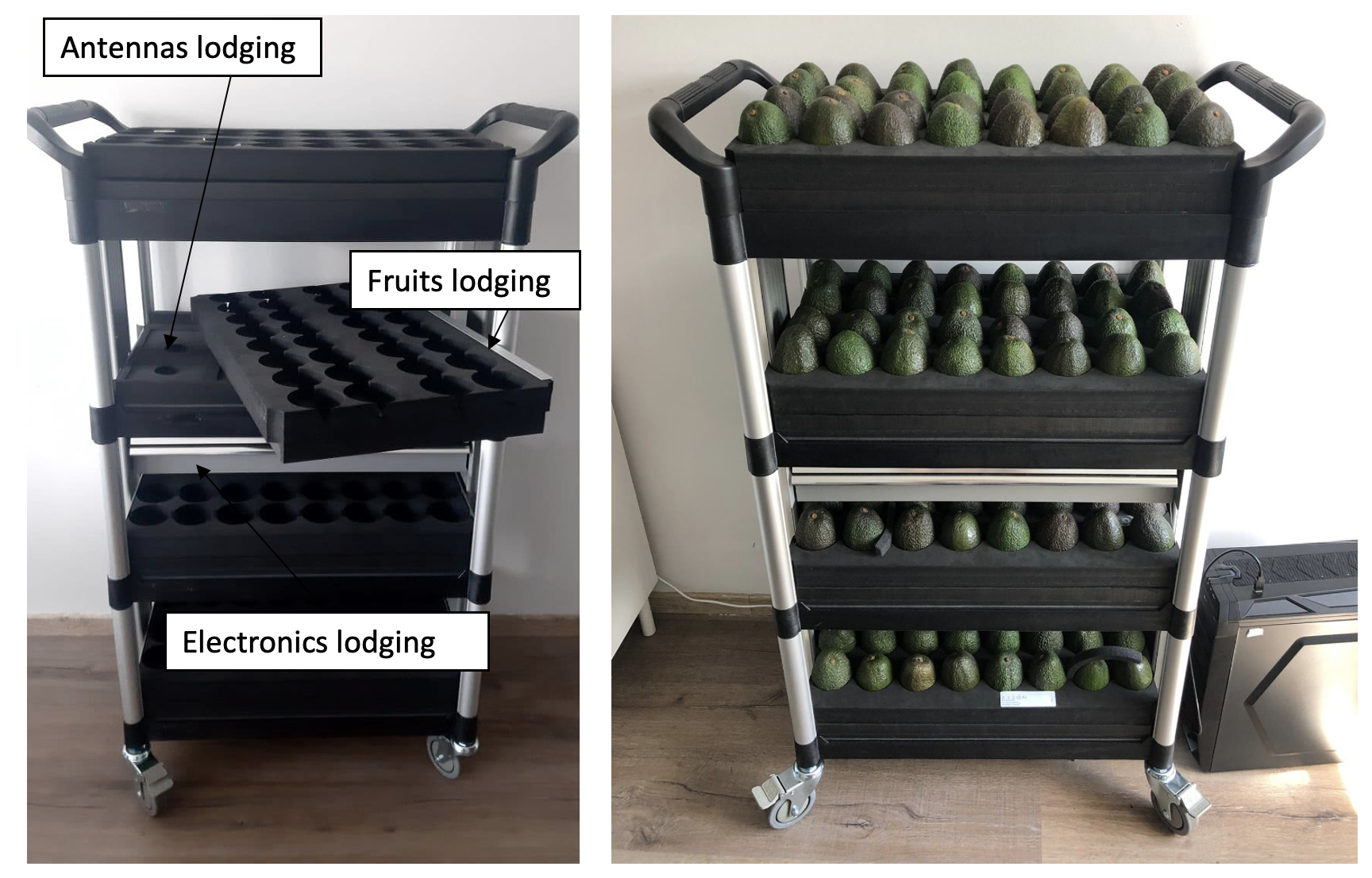}
\caption{Prototype of the smart trolley. Lodging for fruits, antennas and electronics are visible.}
\label{fig:fig_prototype_smart_trolley_lodging}
\end{figure}

The prototype of the trolley is visible in Fig.~\ref{fig:fig_prototype_smart_trolley_lodging}. All the electronic components are placed on an additional, closed shelf. Coaxial cables connecting the antennas are inside the aluminium skeleton. \par 
\begin{figure}[tp]
\centering
\includegraphics[width=8.58cm,height=3.75cm]{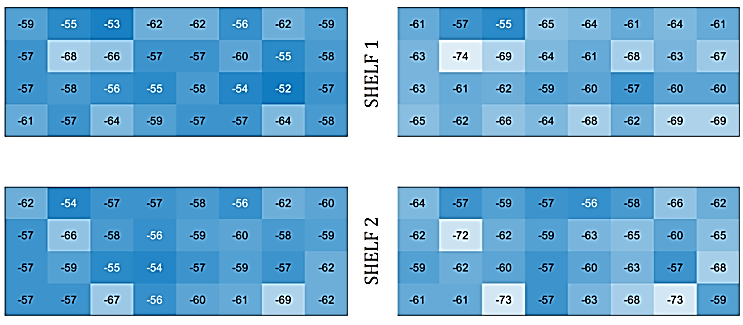} \\
(a) \qquad \qquad \qquad \qquad \qquad \quad (b)
\caption{Measured RSSI for the (a) unloaded and (b) loaded two upper shelves. When the fruit lodging is empty, the average RSSI is almost constant.}
\label{fig:measured_rssi_a_unloaded_and}
\end{figure}

Thanks to the symmetry of the reading architecture and of the arrangement of cables and antennas, the reading coverage is uniform. Fig.~\ref{fig:measured_rssi_a_unloaded_and} shows an example of a coverage map of the top (number~$1$) and second-top (number~$2$) shelf of the trolley for both unloaded (without fruits) and loaded (with fruits) lodgings. The RSSI (averaged value over the three A, B, and C tags) of the unloaded cavities are almost uniform and comprised between $-60$ and $-53$~dBm. The presence of the fruits leads the loaded shelf in operating in different conditions. Due to the typical and uncontrolled variability of the fruits, signals are different. However, communication is always guaranteed, being RSSI typically comprised between $-75$ and $-50$~dBm. \par
\begin{figure}[tp]
\centering
\includegraphics[width=8.57cm,height=3.64cm]{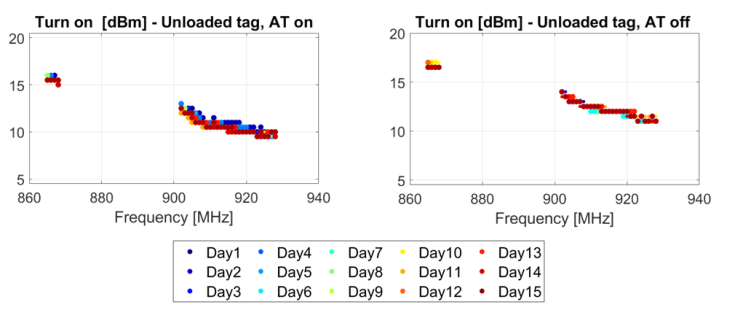} \\
(a) \qquad \qquad \qquad \qquad \qquad \quad (b)
\caption{Measured turn-on power of tag A of an unloaded cavity over $15$ days when (a) the AT is on and (b) when AT is off.}
\label{fig:measured_turnon_power_tag_unloaded}
\end{figure}

\begin{figure}[tp]
\includegraphics[width=8.61cm,height=4.55cm]{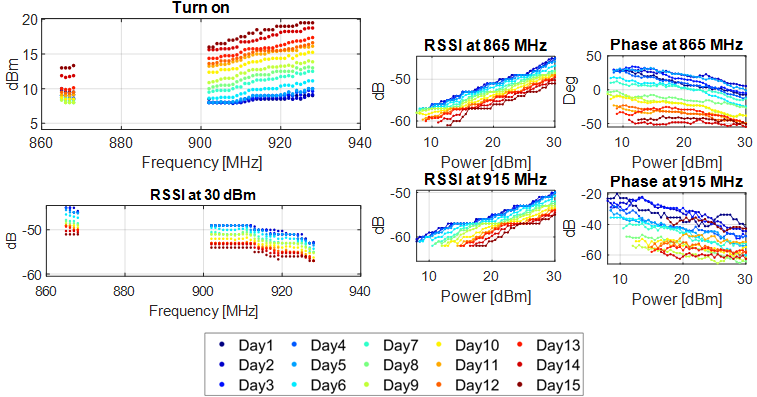}
\caption{ Example of different RF indicators retrieved from a single avocado (tag A) when imposing different interrogation modalities.}
\label{fig:fig_example_different_rf_indicators}
\end{figure}

To evaluate the robustness of the signals against uncontrolled and random phenomena, the trolley was partially loaded by fruits, and the electromagnetic indicators were periodically measured for $15$ days. As an example, Fig.~\ref{fig:measured_turnon_power_tag_unloaded} shows the turn-on power of tag A of an unloaded fruit cavity. Signals are almost constant $\Delta$\( P_{in}^{to}\)$\sim$1 dB, with only typical fluctuations of RFID systems \cite{Amendola15}. Setup can be hence considered robust. Instead, when the fruit is present, RF indicators vary all along the ripening process (an example in Fig.~\ref{fig:fig_example_different_rf_indicators} for the same tag A). The turn-on power is always more than $10$~dB lower than the maximum available power, and therefore the communication link can be deemed robust during the whole ripening process. \par
\section{Classification Algorithm}\label{sec:Classification}
The prediction strategy of the ripening stages relies on machine learning techniques applied to the electromagnetic data collected by the RFID system. Among several possible classifiers, a Support Vector Machine was selected \cite{Scholkopf01} for the easiness of implementation and the speed of classification. SVMs are known as maximum margin classifiers as they find the best separating hyperplane between two classes. Since the recursive application of the algorithm allows the separation of any number of classes \cite{Amendola15}, SVM can be efficiently applied also to the avocado ripening process and to the identification of the four classes in Table~\ref{tab:table_i}. With respect to each value, the algorithm's output can be $0$ if the status of the avocado is evaluated above the threshold or $1$ if the threshold is considered passed. Additional assumptions underlying the construction of the model are:
\begin{enumerate}
	\item the classification model identifies the change of ripening of the fruits with respect to its initial state;
	\item all the fruits are assumed to be unripe at the time zero.
	\item the evolution of the process \(\left\{ C1(t1)\rightarrow C2(t2)\rightarrow C3(t3)\rightarrow C4(t4)\right\}\) with \(t1<t2<t3<t4\) is forced to be monotonic to solve possible uncertainties related to misclassification.
\end{enumerate}
By considering the presence of three tags per fruit, two independent models were investigated:
\begin{enumerate}
	\item \textbf{Tag A model} that considers only the signals from the basal tag A.
	\item \textbf{Tags B-C model} that jointly considers both signals from tags B and C.
\end{enumerate}
The final classification of the ripening state is then obtained by combining via a \textit{logic OR} the outputs coming from the two separate models. The OR port logically implements the physical evolution of the ripening that can start from any point of the fruit. If one of the two classification models estimates the passing of one of the thresholds, the fruit state is then globally considered $1$ with respect to the threshold itself. This feature guarantees the early detection of ripening.\par
The classification algorithm abstains in case of missed readings (tag A and both tags B and C not readable) or discordance on the crossing of the thresholds. \par
The use of SVM involves a three-step procedure, namely \textit{i}) feature selections, \textit{ii}) training, and \textit{iii}) testing. Each step is detailed in the following.
\subsection{Feature Selection}
The training dataset was built by measuring SH and RF indicators of $32$ avocados for seven days by considering four measurement cycles each day. Dataset hence contained $896$ classified fruits, for approximately $880000$ entries. Each fruit was assigned to a ripening class according to the measured SH and the scheme in Table~\ref{tab:table_i}. For reducing fluctuations, a moving average was performed on a $7$-samples window for each fruit, determined through a study on the convergence error \cite{Camera21}. After the filtering, signals were normalized with respect to their initial value. Signals coming from tags B and C were then averaged. \par
\begin{figure}[tp]
\centering
\includegraphics[width=8.58cm,height=8.46cm]{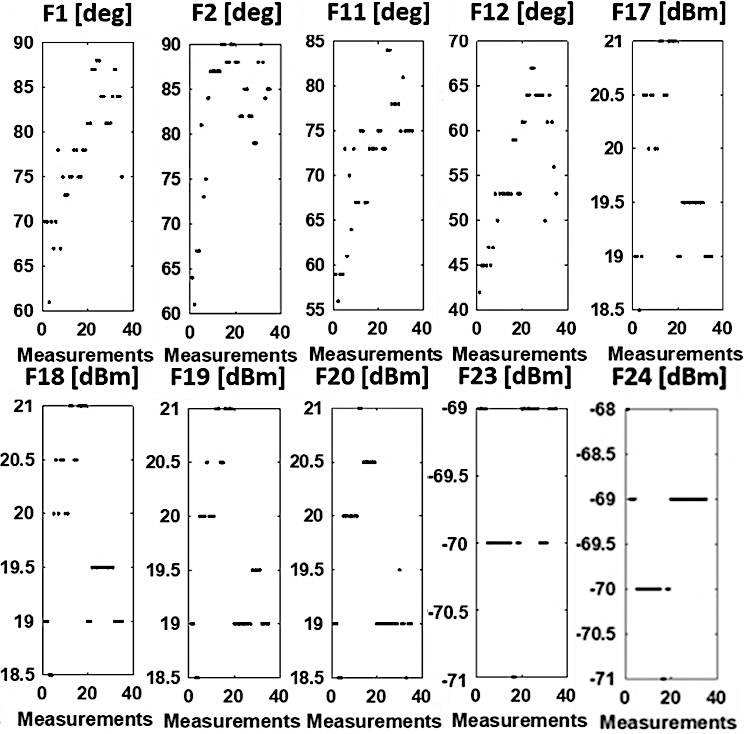}
\caption{An example of ten measured RF features during consecutive measurements.}
\label{fig:example_ten_measured_rf_features}
\end{figure}

The interrogation returns $28$ features. Afterwards, the feature set is thinned based on their Area Under the receiver operating characteristic Curve (AUC) \cite{Fawett16} computed over the training set. In particular, just the top 5 features in the AUC were selected for Tag A model, whereas the top $10$ were chosen for tags B-C model (an example is visible in Fig.~\ref{fig:example_ten_measured_rf_features}, and all the features are listed in Table~\ref{tab:table_2}). The optimal feature set differs for basal and equatorial tags and for each threshold of the normalized SH. This finding highlights the different information content depending on the positions of tags and the ripening stage of the fruit itself. \par
\begin{table*}[tp]
\begin{center}\caption{Selected features (noted by crosses) to feed the classification algorithm depending on the utilized model, frequency, threshold, AT status, and interrogating power}
\includegraphics[width=18cm]{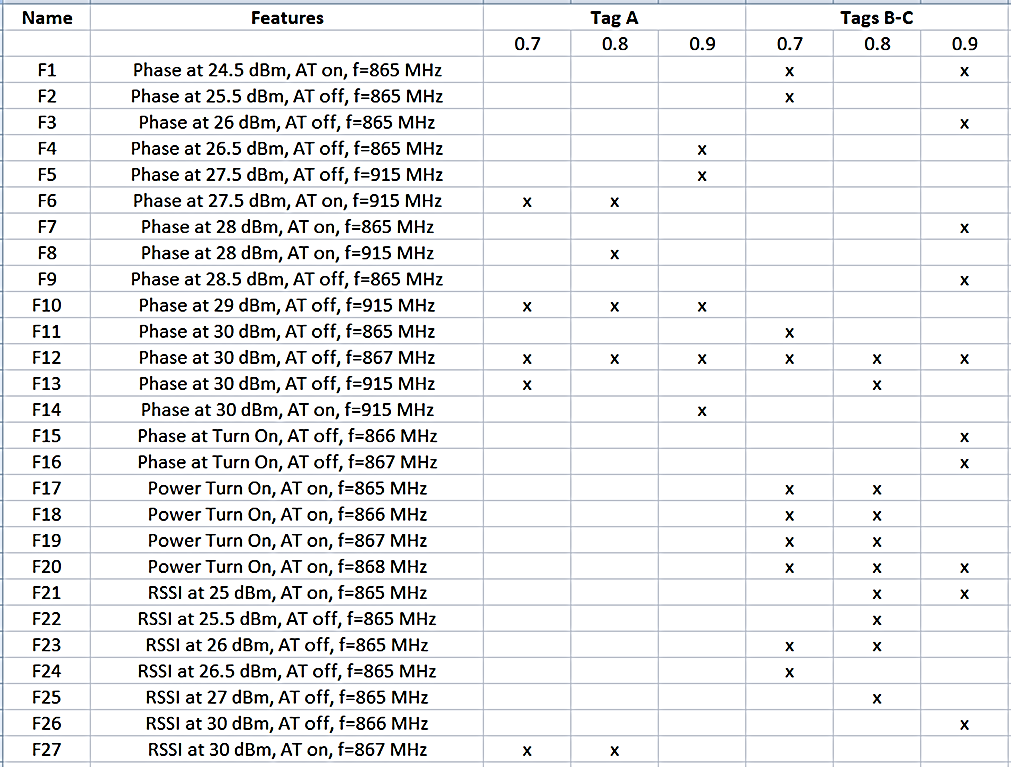}
\label{tab:table_2}
\end{center}
\end{table*}
\subsection{Training and Test}
The selected and preprocessed features are then used to build the two classification models. The kernel is composed of radial basis functions [27], and the kernel scale was set to the inverse of the number of available features. \par
Models are trained and tested by \textit{Leave One Fruit Out} cross-validation \cite{Bishop06}: samples belonging to one fruit were left in the test while the other samples were used for model building. The procedure is then repeated several times by sequential rotation until each fruit has been used for the test. \par
\begin{figure}[tp]
\includegraphics[width=8.58cm,height=4.09cm]{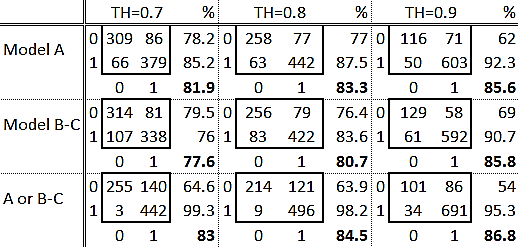}
\caption{Confusion matrices of the three proposed models at different ripeness thresholds obtained on the training set with Leave One Fruit Out cross-validation. In each cell, the number of classifications is indicated.}
\label{fig:fig_confusion_matrices_three_proposed}
\end{figure}

The results are summarized through the confusion matrix (CM) in Fig.~\ref{fig:fig_confusion_matrices_three_proposed}. Each of the $2\times 2$ tables describes the relationship between estimated (column) and actual (row) patterns, i.e., the ability to correctly classify the ripening state \(K=\left\{0,1\right\}\). For each threshold TH, four possible cases can be obtained.
\begin{enumerate}
	\item \textbf{True-negative} (TN). The algorithm correctly evaluates the non-crossing of the threshold (K$=0,0$).
	\item \textbf{False-positive} (FP). The algorithm wrongly evaluates the crossing of the threshold (K$=0,1$).
	\item \textbf{False-negative} (FN). The algorithm wrongly evaluates the non-crossing of the threshold (K$=1,0$).
	\item \textbf{True-positive} (TP). The algorithm correctly evaluates the crossing of the threshold (K$=1,1$).
\end{enumerate}
The first and the last conditions represent hence good evaluations. The more diagonal the matrix is, the more accurate is the evaluation. The accuracy of classification is evaluated as TN/(TN+FP) for row 0, TP/(FN+TP) for row $1$, and (TP+TN)/(TP+TN+FP+FN) for the whole confusion matrix. Results over $840$ evaluations demonstrate an average accuracy ranging between $77\%$ and $86\%$. The logical combination between responses of model A and model B gives better performances, with an average accuracy of $85\%$ for all the thresholds. As is evident by comparing the classification of the A and A OR (B-C) models, model A has the most significant information content, probably because of local processes in the ripening of the fruits. \par
The algorithm did not classify almost $6\%$ of the dataset. This was mainly due to interrogation errors (e.g., missing readings mainly related to two reader antennas not perfectly connected to the cables) and was mitigated in subsequent campaigns by strengthening the hardware of the reading network. \par
\section{Error Analysis and Performance Assessment}\label{sec:Error}
The analysis of the errors and the assessment of the classification performances were performed on the full-loaded, entire trolley. Two experimental campaigns considered unripe fruits that were monitored for seven days at room temperature. Four evaluations were scheduled per day to avoid possible overlapping between different reading sequences over the whole trolley, even in case of delayed missed readings. Each campaign returned approximately $4000$ classifications aside from \(5\%\) of cases of abstention. The first campaign was used for error evaluations, the second one for performance assessment.
\subsection{Analysis of Classification Errors}
\begin{figure}[tp]
\centering
\includegraphics[width=6.18cm,height=4.9cm]{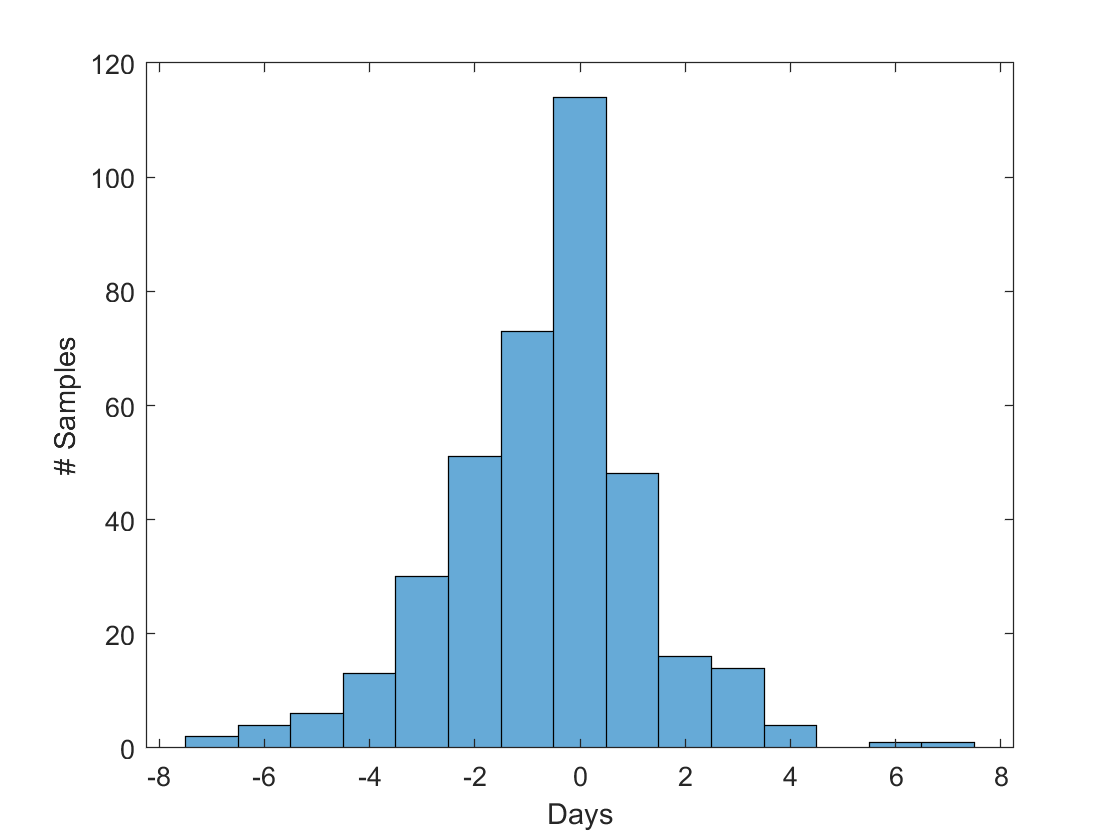}
\caption{Error analysis with respect to the day.}
\label{fig:fig_error_analysis_respect_day}
\end{figure}

A first analysis of the classification errors was related to the "switching day", viz., the day in which the measured SH crosses one threshold TH changing the fruit classification ($K=0\rightarrow1$). Results are visible in Fig.~\ref{fig:fig_error_analysis_respect_day} in terms of cumulative error distributions for all the three values of the threshold (TH$=0.7,$ $0.8$ and $0.9$). The \textit{x}-axis indicates the temporal distance \(D=D_{i}-D_{0}\) in days between the day of the misclassified sample (\(D_{i}\)) and the real switching point (\(D_{0}\)). On the \textit{y}-axis instead, there is the number of errors for each temporal distance. In other words, the performance parameter is the number of misclassification for each value of the difference between the switching day returned by the algorithm \textit{D\textsubscript{i}} and the actual switching day \textit{D\textsubscript{0}} measured by the SH. Regardless of the threshold value, most of the errors occur on the same day of the threshold crossing. Distribution is then quite symmetric, even if the error is slightly higher for negative D, meaning that the algorithm tends to anticipate the switching day and hence the maturation process. Accepting an uncertainty of $\pm1$ a day, the accuracy increases up to $93\%$. \par
\begin{figure}[tp]
\centering
\includegraphics[width=8.61cm,height=4.32cm]{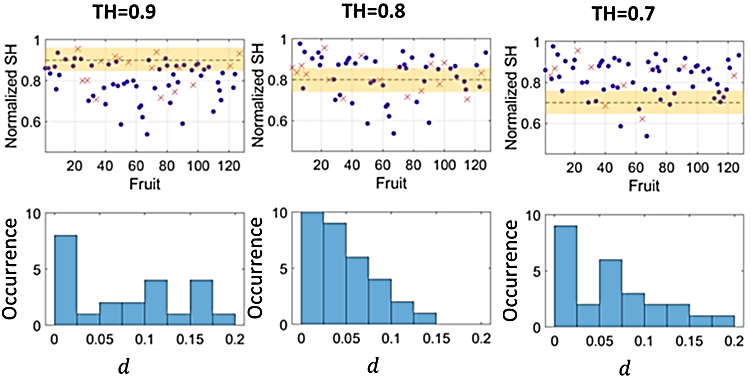} \\
(a) \qquad \qquad \qquad \quad (b) \qquad \qquad \qquad \quad (c)
\caption{Error analysis for each threshold. Fruits are numbered from $1$ to $128$. For each fruit, the red cross means an incorrect classification, while the blue dot is a correct one. (a) TH$=0.9$, (b) TH$=0.8$, and (c) TH$=0.7$. The distribution of the errors is vs \(d =\left(SH_{measured}-TH_{n}\right)\)}
\label{fig:fig_error_analysis_for_each}
\end{figure}

Errors were then analyzed with respect to measured SH. Results are depicted in Fig.~\ref{fig:fig_error_analysis_for_each} for the three thresholds; each marker is an evaluation of the single fruit (univocally identified by a number from $1$ to $128$ on the \textit{x}-axis). On the \textit{y}-axis, there is the measured SH, while the shape of the markers identifies the result of the classification: blue dots represent fruits correctly classified, red crosses represent misclassification with respect to the threshold TH. Errors mainly occur close to the separation hyperplane and, in particular, in the interval \(d =\left(SH_{measured}-TH_{n}\right)\) comprised in the $\pm$5$\%$ of the central value (see the distribution bars in the same Fig.~\ref{fig:fig_error_analysis_for_each}). Errors are more frequent for TH$=0.9$ and TH$=0.8$ because of the uncertainty related to the SH measurements, which are more challenging to be performed when the fruits are harder. By accepting an uncertainty of $\pm5\%$ in the measured SH, a further improvement of about $20\%$ in the accuracy can be evaluated. In this case, the average accuracy observed is $90\%$ over the seven days. \par
It is worth noticing that the measurement of SH through the durometer is typically prone to errors and uncertainties mostly related to the operator. Conventionally, an average over three points is considered, with standard deviations that could reach up to $10\%$ over the whole fruit surface. Consequently, the previous tolerances can be considered acceptable and in line with the typical ripening phenomena. \par
\subsection{Performance Assessment}
\begin{figure}[tp]
\includegraphics[width=8.34cm,height=13.2cm]{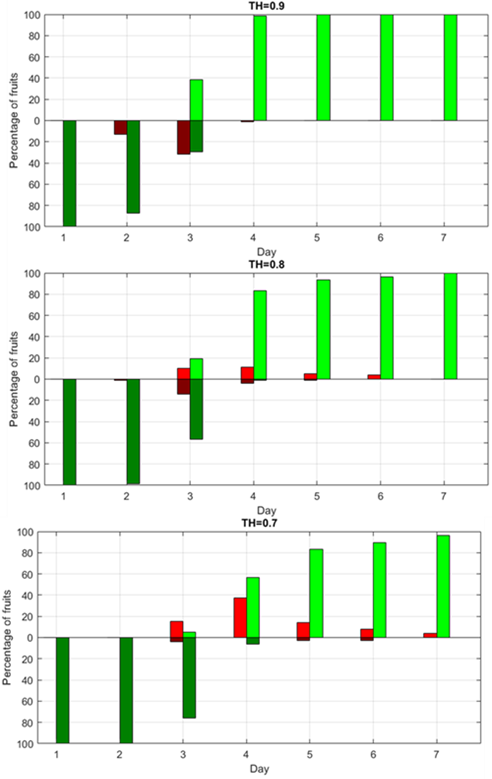}
\includegraphics[width=8.45cm,height=0.43cm]{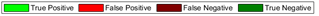}
\caption{Results of the 7-days monitoring of 128 fruits.}
\label{fig:fig_results_7days_monitoring_128}
\end{figure}

Fig.~\ref{fig:fig_results_7days_monitoring_128} shows the classification results of a second, seven-days-long measurement campaign on 128 fruits. \par
Results are classified according to the previous confusion matrix approach. The statistic mode of the four measurements of each day is considered to give a single indication per day. Starting from the unripe state at day $1$ ($100\%$ TN-bars; the algorithm correctly classifies the non-crossing of the threshold of all the fruits), fruits start ripening, hence the height of the TP-bars increases as fruits progressively cross the thresholds TH. Right after the second day of measurement, some avocados reach the SH$=0.9$ value, while the third threshold, SH$=0.7$, is met after $3$-$4$ days. \par
Wrong evaluations mainly occur in the proximity of the maturation, particularly on day $3$ for TH$=0.9$ and day $4$ for TH$=0.7$. Apart from the early threshold of 0.9, the algorithm tends to anticipate the ripening, with the presence of FP evaluations right before the switching days. \par
\begin{figure}[tp]
\centering
\includegraphics[width=8.58cm,height=4.45cm]{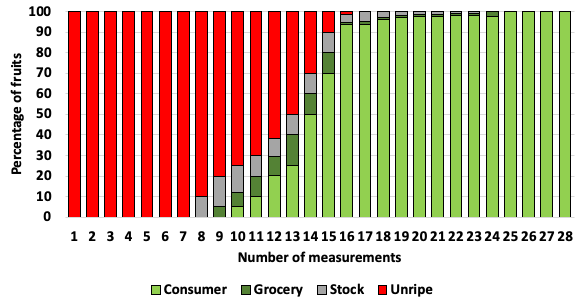}
\caption{Evolution of the classification measurement by measurement of 7days monitoring of 128 fruits. Results of the classifier are reported regardless their correctness.}
\label{fig:fig_evolution_classification_measurement_measurement}
\end{figure}

A summary view of the results of the classification measurement by measurement is visible in Fig.~\ref{fig:fig_evolution_classification_measurement_measurement}, regardless their correctness. The monotonic evolution is clearly distinguishable. In the beginning, all the fruits are reasonably unripe. Along the days, the percentage of ready-to-eat fruits monotonically increases, while the ones of the other two classes decrease. On the fourth day, $80\%$ of the trolley reaches the final state. 
\vspace{1\baselineskip}
\section{Conclusion}\label{sec:Conclusion}
A system for automatically evaluating the ripening of avocados in an industrial scenario has been proposed and tested. Each fruit is monitored by three sensor tags through a multi-level reader network. The adopted architecture enables multi-points ripening monitoring, with augmented detection capabilities in case the ripening starts from different fruit regions. By analyzing RFID signals through SVM, the automatic classification of four ripening stages can be achieved with an average accuracy higher than $85\%$. \par
The high observed variability among fruits further confirms the non-full reliability of the time-based classification methods currently employed \cite{Schaffer13} and the value of the proposed strategy. Compared to the previous works \cite{occhiuzzi2020radio}, the present monitoring platform demonstrated to be capable of retrieving the very early stages of the ripening process (in which the variations of the electromagnetic signals are limited), and hence the trolley can be effectively adopted in the industrial sector, where operators are generally interested in producing fruits ready to be distributed to the consumers rather than \textit{ready-to-be-eaten} by the consumers \cite{occhiuzzi2020radio}. \par
Implementation and usage costs are limited. The entire hardware cost (RF components, mechanical structures, and PC) could be estimated in $\mathbf{3.500,00}$~€ to be further reduced in case of multiple installations. Thanks to the integration of the tags into the trolley, the system can be reused many times with limited maintenance requirements. Furthermore, loading and unloading operations are extremely easy and rapid, thanks to the engineered cavity and the plastic multi-leaf support. Finally, the number of avocados hosted in the trolley is enough to have a representative set of the fruits present in the ripening chamber. Several varieties could be contemporarily monitored, for instance, by differently loading the four shelves. \par
Further improvements are expected by enriching the training dataset with a progressively higher number of measurements. Finally, additional UHF RFID chemical and physical sensors could be integrated into the same infrastructure to better sample and control the environmental condition in the close surrounding of the fruits. \par

\end{document}